\documentclass[a4paper]{jpconf}
\usepackage{graphicx}
\begin{document}
\title{Evidence for $h_{c}$ Production from $\psi^{\prime}$ at CLEO}

\author{Amiran Tomaradze}

\address{Northwestern University, USA. Representing the CLEO Collaboration}

\ead{amiran@bartok.phys.northwesetrn.edu}

\begin{abstract}
Using the $\sim 6$ pb$^{-1}$ of $e^{+}e^{-}$ annihilation data 
taken at $\psi^{\prime}(3686)$ with CLEO III and CLEO-c detectors 
with estimated $\sim 3.0 \times 10^{6}$ $\psi^{\prime}$ events, we have 
searched for the $h_c (1^1P_1)$ state of charmonium in the reaction 
$\psi^{\prime}(3686) \to \pi^{0} h_{c} \to (\gamma \gamma)(\gamma \eta_{c})$. 
The preliminary results are reported.
\end{abstract}

\section{Introduction}

Charmonium spectroscopy has played a crucial role in the understanding of the quark-gluon structure of hadrons and the underlying theory of Quantum Chronodynamics (QCD).  This is primarily due to the fact that the charmonium system is expected to be far less sensitive to the problems associated with relativistic effects and the large value of the strong coupling constant, $\alpha_s$, than the light quark ($u$,$d$,$s$) systems.  Formation cross-sections for charmonium states, their masses and widths are also favorable for precision measurements.  The existing experimental data have defined the spin-independent one-gluon exchange part of the $q\bar{q}$ interaction quite well, however,
the spin dependence of the $q\bar{q}$ potential is not very well understood. 
 In particular, the $\vec{s_1}\cdot\vec{s_2}$ spin--spin, or 
{ hyperfine interaction is not well understood}, because there is 
little experimental data to provide the required constraints for theory.
The primary experimental data required for understanding the $q\bar{q}$ hyperfine  interaction 
is hyperfine, or spin-singlet/spin-triplet splitting:
$\Delta M_{hf}(nL) \equiv \left< M(n^3L_J) \right> - M(n^1L_{J=L})$.

For nearly 20 years, the only hyperfine splitting known was that for the $1S$ states of charmonium,
$\Delta M_{hf}(1S) = M(J/\psi) - M(\eta_c) =  116\pm2 \; \mathrm{MeV.}$
Very recently, Belle, CLEO and BaBar succeeded in identifying 
$\eta_c'(2S)$, 
with the rather surprising result that
$\Delta M_{hf}(2S) = M(\psi') - M(\eta_c') =  48\pm4 \; \mathrm{MeV.}$
Potential model and quenched lattice calculations predicted 
a larger $\Delta M_{hf}(2S)$ [1].  

It is of great importance to find out how the hyperfine 
interaction manifests itself in $P$ states, i.e., to find
$\Delta M_{hf}(1P)\equiv M(< ^{3}P_{J} >)-M(^{1}P_{1})$.
With scalar confinement, $\Delta M_{hf}(1P)=0$ is expected.  
It is necessary to determine if this is true. The c.o.g. of $^3P$ states, $M(< ^{3}P_{J} >)$, 
is well measured,  $M(< ^{3}P_{J} >)$=3525.3$\pm$0.1 MeV.
What is needed is to identify $h_{c}$ and make a precision measurement
of its mass.

\section{Prior Experimental Searches for $h_c$}

The Crystal Ball experiment at SLAC made a search for $h_c$ in 1982 [2].
The search was unsuccessful and they
reported 95\% confidence limits of 
$B(\psi' \to \pi^0 h_c \: , \: h_c \to \gamma \eta_c) <$ 0.32\%
in the range  $M_{h_c} = 3440 - 3543$ MeV. The next 
search for $h_c$ was made by the Fermilab experiment E760 [3] in 
the reaction 
$p\bar{p} \to h_c \to \pi^0 J/\psi$.
It was claimed that a statistically significant enhancement was observed and that the data indicated 
$M(h_c) = 3526.2\pm 0.15\pm 0.2$ MeV.
However, such an enhancement has not been
confirmed by the successor Fermilab E835 experiment, with significantly
higher statistics [4,5].
The E835 experiment also  searched for $h_c$ in the reaction    
$p\bar{p} \to h_c \to \gamma \eta_{c} $. Preliminary evidence 
at the $\sim$3$\sigma$ significance level has been recently reported with  
$M(h_c) = 3525.8\pm 0.2\pm 0.2$ MeV [5]. No positive evidence
has been reported yet by Belle and BaBar Collaborations.

 It is fair to say that at present there is no convincing experimental evidence for $h_c$ observation.  

\section{CLEO Searches and Results}

The above considerations have motivated us to search for $h_c$ in the $\sim 6 pb^{-1}$ data taken at CLEO with estimated $\sim 3.0 \times 10^{6}$ 
$\psi^{\prime}$ events, in the reaction

$$ \psi' \to \pi^0 h_c \; , \; h_c\to\gamma \eta_c.$$

We search for this channel: (a) without using $\eta_{c}$ decays
   (INCLUSIVE approach, see Section 3.1), and (b) using six dominant $\eta_{c}$ decay modes
   (EXCLUSIVE approach, see Section 3.2).
In both methods we search for $h_{c}$ in the mass recoiling against
$\pi^{0}$ from decay $\psi^{\prime} \to \pi^{0} h_{c}$.
This method benefits from the excellent resolution of the CLEO calorimeter.

\subsection{Inclusive Analyses}

Two independent analyses have been performed, and results from the 
two are consistent. I will describe
one of them in detail, and will later mention the differences
between the two analyses. 
We use the following selection criteria: $N_{shower}\ge$3,
$N_{track}\ge$2.
The selection of the showers and charged particles 
are done using the standard CLEO quality cuts. 

We reconstruct $\pi^{0}$'s by requiring that the two photon 
invariant mass be in the range $M_{\gamma\gamma}$=135$\pm$15 MeV,
and that the two photons have been succesfully fitted to  $\pi^0$.
We require that there be only 
one $\pi^{0}$ in the event with a recoil mass in the expected 
$h_{c}$ mass range of 3526$\pm$30 MeV.

The $\psi^{\prime} \to \pi^{+}\pi^{-} J/\psi$ and 
$\psi^{\prime} \to \pi^{0}\pi^{0} J/\psi$  events are 
removed  by cutting on the recoil mass of $\pi^{+}\pi^{-}$  and
$\pi^{0}\pi^{0}$, respectively.

We define hard $\gamma$'s, the possible candidates from  
$h_{c} \to \gamma \eta_{c}$ decays, by $E_\gamma > 400$ MeV.  
We reject such $\gamma$'s which make a $\pi^{0}$  
or $\eta$ with any other $\gamma$'s.  We then require that the 
energy of hard $\gamma$ should be in the range 
$E_{\gamma}$=503$\pm$40 MeV. 

The background in data has been fitted in three ways: (a) ARGUS shape, 
$y=x\times\sqrt{1-(x/a)^{2}}\times exp(b\times[1-(x/a)^{2}])$,
(b) second--order polynomial shape, (c) background shape from Monte Carlo.
The significance levels are obtained as 
$\sigma\equiv\sqrt{-2\ln(L_0/L_{max})}$, where $L_{max}$ and  $L_0$ are 
the  likelihoods of the fits with and without the $h_c$ resonance.

The analysis on the Monte Carlo samples has been performed.
The event selection criteria applied to the Monte Carlo samples
were identical to those applied to the data. 10,000 signal Monte Carlo 
events for the channel 
$\psi^{\prime} \to \pi^{0} h_c \to (\gamma\gamma) (\gamma\eta_c)$
were simulated. The recoil mass distribution against $\pi^{0}$  
in signal Monte Carlo, for input $\Gamma(h_{c})$=0 MeV is well
fitted with a double Gaussian with parameters $\sigma_{1}$=1.3 MeV, 
$\sigma_{2}$=3.7 MeV, and the fraction of second Gaussian was 0.43. 
These parameters, which represent the $\pi^{0}$ recoil mass 
resolution at $h_{c}$, are used to fit the signal in the data. 
The selection efficiency was about 16\%.
We also analyzed a sample of $\sim12\times 10^6$ generic $\psi'$ 
Monte Carlo events 
(events containing all measured $\psi'$ decays except those via $h_c$)
in four separate samples, each with approximately the same size ($\sim3\times 10^6$) as the data. The signal Monte Carlo events were added in 
to the generic Monte Carlo.
The study of these Monte Carlo events  yielded good agreement
between input and output values for both, $M(h_{c})$ and
 $B(\psi^{\prime} \to \pi^{0} h_c) \times B(h_c \to \gamma\eta_c)$,
and showed that the analysis is sensitive to $h_{c}$ production.

Figure 1  shows recoil mass distribution against $\pi^{0}$ in data.
The results of the fit are: $M(h_c)=3524.4\pm 0.7$ MeV, 
N$(h_c$) = $156\pm48$, significance($h_c$) = 3.3 $\sigma$.

An independent alternative analysis has been done. 
The main difference is that in this analysis instead of constraining 
the energy of the hard photon, the  constraint is put 
in terms of recoil against $\pi^{0}\gamma$ ($\eta_{c}$ mass).
The results are consistent with those shown above. 
Thus our preliminary CLEO results from two inclusive analyses are:\\
$\bullet$ $M(h_{c})$=3524.8$\pm$0.7(stat)$\pm\sim$1(syst) MeV, \\
$\bullet$ $B(\psi^{\prime} \to \pi^{0} h_c) \times B(h_c \to \gamma\eta_c)$ 
=(2--6)$\times 10^{-4}$, \\
$\bullet$ The significance of $h_{c}$ detection $>$ 3 $\sigma$.

Estimates of systematic errors in $M(h_{c})$ have been made by 
studying the following: $\pi^{0}$ energy scale, background shapes,
Monte Carlo input/output differences, non-resonant background,
assumed $h_{c}$ width, binning effects, cut variations, and finally,
the difference in $M(h_{c})$ in the two inclusive analyses.

\begin{figure}[h]
\includegraphics[width=18pc]{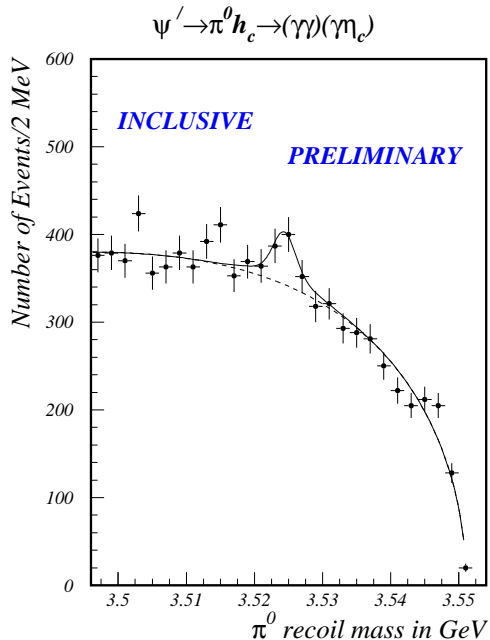}\hspace{2pc}%
\begin{minipage}[b]{18pc}\caption{\label{label}
Distribution of the recoiling mass against $\pi^{0}$ in 
data (inclusive analysis). 
The curves are the results of the fit.
The shape of the signal is assumed as Double Gaussian, and the shape of the 
background is assumed as ARGUS shape (see text).}
\end{minipage}
\end{figure}

\subsection{Exclusive Analysis}

Six $\eta_{c}$ decay modes which have reasonably high PDG04 branching
ratios have been studied:
$K_{s}K^{\pm}\pi^{\mp}$,
$K^{+}K^{-}\pi^{0}$, 
$K^{+}K^{-}\pi^{+}\pi^{-}$,
$2\pi^{+}2\pi^{-}$, 
$\pi^{+}\pi^{-}\eta$ ($\eta \to \gamma\gamma$), and
$\pi^{+}\pi^{-}\eta$ ($\eta \to \pi^{+}\pi^{-}\pi^{0}$).

Standard CLEO selections are used for showers, tracks, and particle
identification.
The total energy--momentum conservation  of the event has been required,
and the invariant mass of the $\eta_c$ decay candidates  
are required to be close to the nominal $\eta_c$ mass (within 50 MeV).
Figure 2(upper plot) shows the  $\pi^0$ recoil mass distribution
for the sum of the six exclusive channels. The fit results are:\\ 
$\bullet$ $M(h_{c})$=3524.4$\pm$0.9(stat) MeV, \\
$\bullet$ N$(h_{c})$=15.0$\pm$4.2, \\
$\bullet$ The significance of $h_{c}$ detection  $\sim$~5~$\sigma$.\\
Note that the significance is calculated using likelihood differences.
The background estimation by using $\eta_{c}$ 
sidebands(closed circles in Figure 2, lower plot), or by using
generic Monte Carlo events(open squares in Figure 2, lower plot), 
yield consistent results. 
No estimate of the systematic uncertainty in $M(h_{c})$ has been made so far. 

\begin{figure}[h]
\begin{center}
\rotatebox{270}{\includegraphics[width=2.5in]{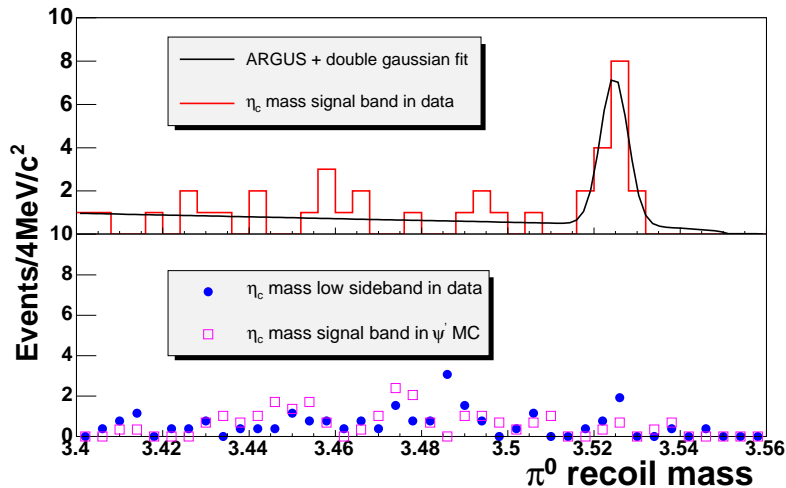}}
\caption{\label{label}
Distribution of the recoiling mass against $\pi^{0}$ (exclusive analysis).}
\end{center}
\end{figure}

\section{Summary}

We have analyzed $\sim 3.0 \times 10^{6}$ $\psi^{\prime}$ from
CLEO III and CLEO-c  
to search for $h_{c}(^1P_{1})$ production in the reaction 
$\psi^{\prime} \to \pi^{0} h_{c}$, $h_{c} \to \gamma \eta_{c}$ 
by two methods 1. INCLUSIVE -- which does not use $\eta_{c}$ decay modes, 
2. EXCLUSIVE -- which uses six hadronic decay modes of $\eta_{c}$.\\
In the recoil mass spectrum of $\pi^0$, we see an 
enhancement in both analyses.\\
$\bullet$ In the inclusive analysis we obtain

$M(h_{c})$=3524.8$\pm$0.7(stat)$\pm\sim$1(syst) MeV,

$B(\psi^{\prime} \to \pi^{0} h_c) \times B(h_c \to \gamma\eta_c)$ 
=(2--6)$\times 10^{-4}$, 

significance  of $h_{c}$ detection $>$3 $\sigma$. 

Thus, $\Delta M_{hf} \equiv \left< M(\chi_J) \right> - M(^1P_1)$ = 0.5$\pm$0.7(stat)$\pm\sim$1(syst) MeV.\\
$\bullet$ In the exclusive analysis we obtain

 $M(h_{c})$=3524.4$\pm$0.9(stat) MeV,  

 significance  of $h_{c}$ detection $\sim$ 5 $\sigma$. \\
$\bullet$ The inclusive and exclusive results for $M(h_c)$ are in excellent agreement.

\section*{Acknowledgments}
I would like to thank Kam Seth, Sean Dobbs, Zaza Metreveli (Northwestern
University), Jon Rosner (University of Chicago), Hajime Muramatsu 
(Syracuse University), Datao Gong, Yuichi Kubota (University of Minnesota),
for their contributions to this analysis. I thank David Asner,
Gocha Tatishvili, Hanna Mahlke-Kruger, Helmut Vogel, Roy Briere, Steven Dytman,
Todd Pedlar, for their helpful comments.
Special thanks goes to Rich Galik for his useful comments and 
continuous help during preparation of these results.

This work was supported by U.S. Department of Energy.

\smallskip

\end{document}